\documentclass[prd,aps,floatfix,nofootinbib,preprint ,tightenlines,superscriptaddress]{revtex4}
\usepackage{dcolumn}% Align table columns on decimal point
\usepackage{amsmath}
\usepackage{latexsym}
\usepackage{graphicx}
\usepackage{bm}

\newcommand{\bee}{\begin{equation}}
\newcommand{\ee}{\end{equation}}
\newcommand{\beea}{\begin{eqnarray}}
\newcommand{\eea}{\end{eqnarray}}

\begin{document}
%%%%%%%%%%%%%%%%%%%%%%%%%%%%%%%%%%%%%%%%%%%%%%%%%%%%%%%%%%%%%%%%%%%%%%
\title{Curve collapse for the isospin - 2 pion scattering length from QCD with 3, 4, and 5 colors}
\author{Thomas DeGrand}
\affiliation{Department of Physics,
University of Colorado, Boulder, CO 80309, USA}
\email{thomas.degrand@colorado.edu}

\date{\today}

\begin{abstract}
I show comparisons of the pion  - pion scattering length in the isospin - two 
channel from simulations of of QCD with $N_c = 3$, 4 and 5 colors
and $N_f = 2$ flavors of degenerate mass fermions.
The scattering length varies as $1/N_c$, as expected from large $N_c$ counting arguments.
\end{abstract}
\maketitle

\section{Introduction}

The interaction strength of mesons of QCD with a large number of colors  $N_c$ is expected to decrease
as  $N_c$ is taken to large values: the four-point scattering amplitude is expected to fall as $1/N_c$ \cite{tHooft:1973alw,tHooft:1974pnl,Witten:1979kh}.
Hadronic interactions, at least at low energies, are nonperturbative and so it would be interesting to have
nonperturbative checks of this result.

This is (most likely) easiest to observe via the interaction of pions; specifically in the
isospin-2 ($I=2$) channel. A measurement of the strength of  interaction,
parameterized via the scattering length $a_0^{I=2}$, was recently carried out 
in a  simulation of QCD with $N_f=4$ flavors of degenerate fermions and $N_c=3-6$ colors,
and the appropriate scaling was observed \cite{Baeza-Ballesteros:2022azb}.

Recently, I participated in a comparison of chiral perturbation theory to a
 lattice calculation of
the pseudoscalar mass and decay constant of
QCD with $N_c=3$, 4 and 5 colors and $N_f=2$ flavors of degenerate mass fermions 
 \cite{DeGrand:2023hzz}. It is easy to
repurpose the data sets to compute  $I=2$ scattering lengths and compare them across $N_c$,
although the results will not be as clean as one would expect for a dedicated study.
 That is the subject of this little note.

Calculations of scattering lengths in $N_c=3$ QCD date back over thirty years and the technology for
doing these calculations has become quite involved. The analysis done here is very naive and would not
be state of the art for $N_c=3$ in terms of lattice volumes, fermion masses, or methodology for propagators.
But none of that is really needed to observe the expected $a_0^{I=2} \propto 1/N_c$ behavior: the ``curve collapse''
in the title. Large $N_c$ counting comes from an analysis of the color weight of Feynman diagrams (or maybe better to say,
from the color weight associated with topological classes of amplitudes) and comparisons across $N_c$ could be done
 in finite simulation volume, at any set of quark masses, or away from the continuum limit. 

The quantity which in a lattice simulation gives a scattering length is the energy shift of a  two particle state in 
a box of linear size $L$ and volume $L^3$. The relation was given long ago by L\"uscher \cite{Luscher:1986pf},
\bee
\Delta E = E_2 - 2m = -\frac{4\pi a_0}{mL^3} [1 + c_1\frac{a_0}{L} + c_2 \frac{a_0^2}{L^2} ]+ O(L^{-6}).
\label{eq:scattlength}
\ee
$E_2$ is the energy of the two particle state, $m$ is the energy of a single particle.
Fractional corrections to $\Delta E \propto L^{-3}$ involve two constants $c_1 = -2.837297$ and $c_2=6.375183$.
The quantity $a_0$ is the scattering length, in a convention where it is related to the s-wave 
scattering phase shift $\delta_0$ at
center of mass momentum $p$ via
\bee
\frac{1}{a_0} = p \cot \delta_0 + O(p^2).
\ee

Without further ado, we continue: Sec.~\ref{sec:method} briefly describes the lattice calculation. Results
are found in Sec.~\ref{sec:results}.  It is easy to see that $a_0^{I=2} \propto 1/N_c$ over the range of pion masses kept in the study.
Sec.~\ref{sec:chipt}  is a too-long discussion of my attempts to do fits of my data to the expectations of chiral perturbation theory.
A few conclusions are made in Sec.~\ref{sec:conc}.
Impatient readers uninterested in details should jump immediately to look at Fig.~\ref{fig:ncmpia0vst0mpi2},
which summarizes the results of this study.

\section{Methodology\label{sec:method}}
\subsection{Generic remarks}
A long description of the action, simulation methodology and  analysis
for the data sets used in this project is to be found   in Ref.~\cite{DeGrand:2023hzz}. Here is a quick summary:

The gauge action  is the
 Wilson plaquette action, with the bare gauge
coupling $g_0$ parameterized by  $\beta = 2N_c / g_0^2$.
Two flavors of degenerate mass  Wilson-clover fermions are simulated.
Configurations are generated using
 the Hybrid Monte Carlo (HMC)  algorithm \cite{Duane:1986iw,Duane:1985hz,Gottlieb:1987mq}
with a multi-level Omelyan integrator \cite{Takaishi:2005tz} and
multiple integration time steps \cite{Urbach:2005ji}
with one level of mass preconditioning for the fermions \cite{Hasenbusch:2001ne}.

The fermion action uses normalized hypercubic
 smeared links~\cite{Hasenfratz:2001hp,Hasenfratz:2007rf,DeGrand:2012qa} as gauge connections.
Simulations use the
arbitrary $N_c$ implementation of Ref.~\cite{DeGrand:2016pur}.
The action is written in terms of the hopping parameter $\kappa=(2m_0^q a+8)^{-1}$
rather than the bare quark mass $m_0^q$ and the lattice spacing $a$.
The clover coefficient is fixed to its tree level value, $c_{\text{SW}}=1$.
 The gauge fields obey periodic
boundary conditions; the fermions are periodic in space and antiperiodic in time.
Lattice volumes are a mix of $L^3\times T = 16^3\times 32$ and $24^3\times 32$  sites.
 The lattice sizes were chosen to minimize finite volume effects
for the calculations of Ref.~\cite{DeGrand:2016pur}. For the present
calculation, where a volume-dependent energy difference is used to extract an observed quantity,
this is a defect; ideally, one would combine simulations from many volumes at each bare parameter set
in an extraction of $\Delta E$.

 Hadron correlators are
constructed from  fermion propagators in Coulomb gauge, with Gaussian sources and
 $\vec p=0$ point sinks.
 All results come from a standard full correlated
analysis involving fits to a wide range of $t$'s.
Best fits are chosen with the ``model averaging'' ansatz of Jay and Neil \cite{Jay:2020jkz}.

Recent discussion of expectations for $a_0^{I=2}$ from chiral perturbation theory
(see the  Flavour Lattice Averaging Group (FLAG)
 review \cite{FlavourLatticeAveragingGroupFLAG:2021npn}) is done in terms of
the ratio of pseudoscalar mass and decay constant at nonzero fermion mass,
\bee
\xi = \frac{m_{PS}^2}{8\pi^2 f_{PS}^2}.
\label{eq:xi}
\ee
The  pseudoscalar decay constant is defined through
\bee
\langle 0| \bar u \gamma_0 \gamma_5 d |\pi\rangle = m_{PS} f_{PS}.
\label{eq:fpi}
\ee
With this definition, $f_{\pi} \sim 132$ MeV in QCD. 
The continuum decay constant is related to the lattice one by  $f_{PS}= 2\kappa Z_a f_{PS}^{lattice}$
where $Z_A$ is a scheme matching factor, computed in the
``regularization independent'' or RI scheme \cite{Martinelli:1994ty}.

When needed, the lattice spacing is set via the flow 
parameter $t_0$ \cite{other,Luscher:2010iy}.
To give some context to the results, $t_0/a^2$ ranges from about 1.0 to 2.3 for $N_c=3$,
from about 1.1 to 3.3 for $N_c=4$, and from about 1.5 to 3.5 for $N_c=5$.
The nominal value for the  $N_f=2$ flow parameter is
$\sqrt{t_0}=0.15$ fm from Ref.~\cite{Sommer:2014mea}.

\subsection{Methodology  specific to this project}

\begin{figure}
\begin{center}
\includegraphics[width=0.8\textwidth,clip]{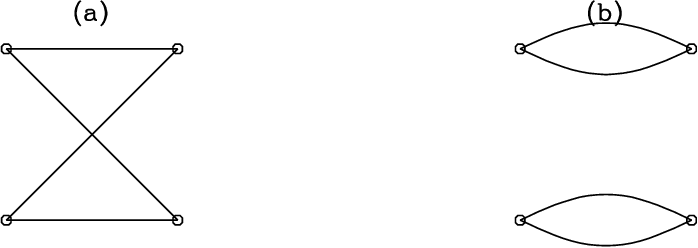}
\end{center}
\caption{ The two Wick contractions whose difference gives the $I=2$ correlator:  (a) the connected correlator, $C_C(t)$
and (b) the disconnected one, $C_D(t)$.
\label{fig:graph}}
\end{figure}

The $I=2$ correlator is the difference of two Wick contraction terms shown in Fig.~\ref{fig:graph},
\bee
C_R(t)= C_C(t)-C_D(t).
\ee
$C_D(t)$ is just the square of the single-pion correlator $C(t)$.
$C_R(t)$ contains a constant term inherited from $C_D(t)$  where the two components
 wrap around the $t$ axis in opposite directions, so (asymptotically in $t$)
\bee
C_R(t)=A\cosh(E_2(t-T/2)) + B
\ee
where  $E_2$ is the energy of the propagating two-particle state. 
Getting a mass difference from separate fits
to $C(t)$ and $C_R(t)$, even folded into a jackknife, produces a noisy result.
The literature devoted to calculation of scattering lengths, or, more 
generally, the literature devoted
to the calculation of energies of many particle states in finite volume, shows  the way to go.
I will follow  Ref.~\cite{Umeda:2007hy} and extract $\Delta E$ from the ratio
\bee
R(t) = \frac{ C_R(t+1)-C_R(t-1)}{C(t+1)^2 - C(t-1)^2}
\label{eq:rrr1}
\ee
whose asymptotic behavior is
\bee
R(t) = R_0 [\cosh(\Delta E\tau) + \sinh(\Delta E \tau) \coth(2m_1 \tau)]
\label{eq:rrr2}
\ee
where  $m_1$ is the mass of the pion, $\Delta E=E_2-2m_1$,
$\tau=t-T/2$ and $R_0$ is a $\tau-$independent constant.
Both $\Delta E$ and its uncertainty are determined
via a a jackknife average: term by term, a fit to $C(t)$ is used to produce produce a model-averaged $m_1$
which is input to Eq.~\ref{eq:rrr2} as it is fit to the ratio of correlators.

The ratio $a_0/L$ can be as large as 0.1 for the $L=16$ $SU(3)$ data sets.
This means that the correction terms in Eq.~\ref{eq:scattlength} are not negligible.
So I use the whole formula, Eq.~\ref{eq:scattlength}, to extract scattering lengths from $\Delta E$
which are accurate to $O(L^{-5})$.
Formulas going to even higher order in $a_0/L$ can be found in the literature
\cite{Beane:2007qr,Hansen:2015zta,Romero-Lopez:2020rdq}. I implemented them and saw that while the
$O(a_0/L)$ correction term is important, the effects of higher order terms
are negligable.

The combination  $m_{PS}a_0^{I=2}$ will be needed. Its values and uncertainties are also computed by jackknife.

\section{Results\label{sec:results}}
\begin{figure}
\begin{center}
\includegraphics[width=0.8\textwidth,clip]{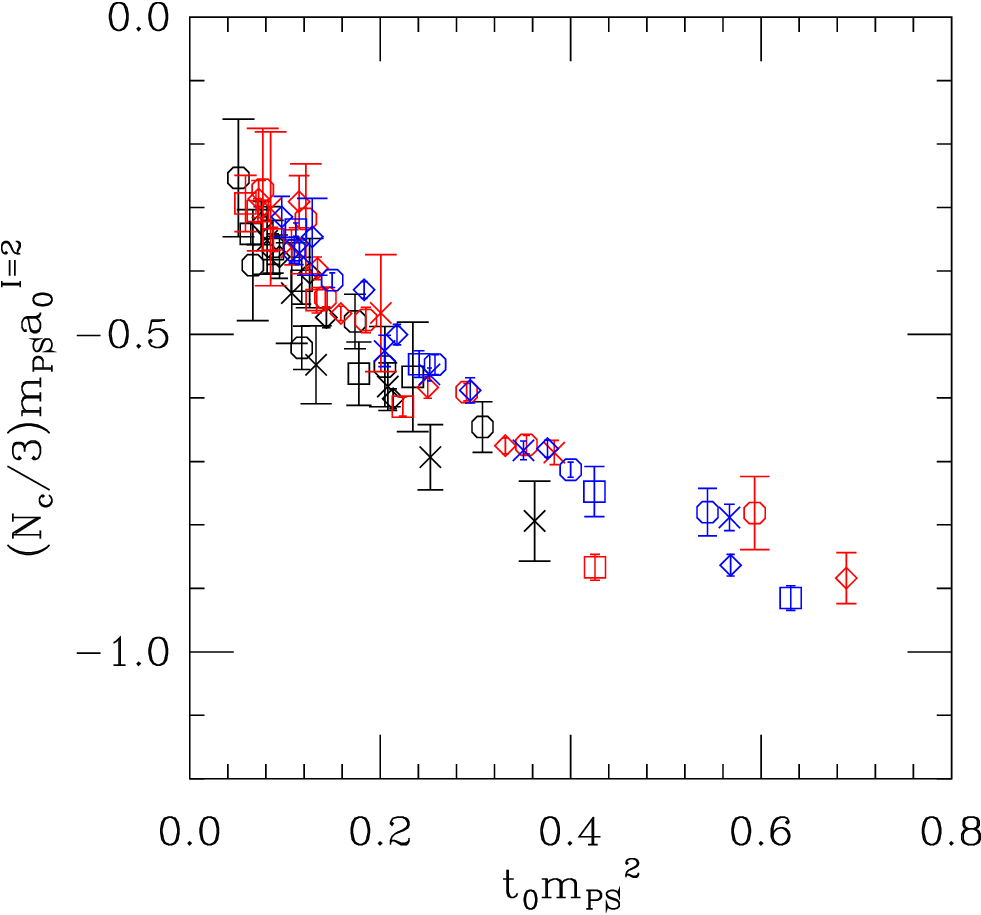}
\end{center}
\caption{ $(N_c/3)m_{PS}a_0^{I=2}$ versus  $t_0 m_{PS}^2$. Black points are
$N_c=3$, red ones are $N_c=4$ and blue ones are $N_c=5$. Different plotting symbols label
different beta values and are listed in the text.
\label{fig:ncmpia0vst0mpi2}}
\end{figure}

Results are collected in Tables \ref{tab:su3tab}-\ref{tab:su5tab}.
The most interesting (to me) result of my study is a  plot of the scattering length versus
(squared) pseudoscalar
 mass, scaled to show curve collapse. Color counting from
 Fig.~\ref{fig:graph} indicates that $a_0$ should scale 
as $1/N_c$ (with corrections which can be represented as a series in higher powers of $1/N_c$), and we already know
that the dependence of meson masses on fermion masses is nearly $N_c$ independent, so a plot of
 $(N_c/3)m_{PS}a_0^{I=2}$ versus  $t_0 m_{PS}^2$ should show curve collapse across $N_c$.
This  is shown in Fig.~\ref{fig:ncmpia0vst0mpi2}.
The $y-$ axis is taken to be
 $m_{PS} a_0^{I=2}$ since that is the combination most often presented in the $N_c=3$ lattice literature.
Different $N_c$ values are represented by different colors, black for $N_c=3$, red for $N_c=4$, blue for $N_c=5$.
Different bare gauge couplings, and hence different lattice spacings, are represented by
different plotting symbols.
In all cases the ordering of symbols (squares, diamonds, octagons, crosses) corresponds to an ordering in decreasing
lattice spacing.
The symbols are
\begin{itemize}
\item For $N_c=3$,
 squares for $\beta=5.25$,
diamonds for $\beta=5.3$,
octagons for $\beta=5.35$,
crosses for $\beta=5.4$
\item For $N_c=4$,
 squares for $\beta=10.0$,
diamonds for $\beta=10.1$,
octagons for $\beta=10.2$,
crosses for $\beta=10.3$
\item For $N_c=5$
 squares for $\beta=16.2$,
diamonds for $\beta=16.3$,
octagons for $\beta=16.4$,
crosses for $\beta=16.6$.
\end{itemize}

Fig.~\ref{fig:ncmpia0vst0mpi2}
shows that, once again, naive expectations for large $N_c$ counting are met: curve collapse for $N_c a_0^{I=2}$ says that 
the strength of interactions between pseudoscalar mesons in the $I=2$ channel decreases with $N_c$ as $1/N_c$
(or better to say, that the amplitude for $\pi^+ \pi^+ \rightarrow \pi^+ \pi^+$ at threshold 
decreases like $1/N_c$). This is a complete result on its own: large $N_c$ comparisons 
can be done at any value of the lattice spacing or any value of the fermion masses.

\section{Attempts at fits to chiral perturbation theory\label{sec:chipt}}
It is customary in lattice calculations of pion scattering lengths in QCD to make comparisons to the expectations
of chiral  perturbation theory. Typically, the predictions are to the
 one loop (or next to leading order or NLO) formula.
I attempted to do this with my data; the attempts were not particularly successful. 
This section  is a bit too long, but it might possibly be useful for anyone who might revisit this topic.

In 2011 Bijnens and Lu \cite{Bijnens:2011fm} carried out the full two loop
(or next to next to leading order or NNLO) calculation of pion scattering lengths.
(Note that their convention for the scattering length 
is $m_{PS}/a_0 = p \cot \delta_0 + O(p^2)$,  so their $a_0(BL)$ is equal
to $m_{PS}a_0$ here. Note also  that they label $a_0^{I=2}$ as  $a_0^{SS}$.)
 They have a plot, their Fig.~5,
of the lowest order (LO), next to leading order (NLO) and 
next to next to leading order (NNLO) results for $m_{PS}a_0^{I=2}$, for reasonable choices
 for the low energy constants, for $N_f$ =2 and $N_c=3$.
Converting the picture's $M_{phys}^2$ (their
notation) to my $t_0 m_{PS}^2$, using a nominal $\sqrt{t_0}=0.15$ fm, one sees that essentially all my data points
lie in a regime in which the difference between NNLO and LO contributions is about twice as big as the
change in prediction from LO to NLO. 

One cannot help feeling that once NNLO effects are as big as NLO effects, there is not much point in 
trying to do an NNLO fit, since we are probably living outside the domain of convergence of the chiral expansion.
Nevertheless, let us press on. The best that one can hope for is that the
situation is similar to what was encountered in chiral fits to the pseudoscalar mass and decay constant
in Ref.~\cite{DeGrand:2023hzz}. There, it was possible to do  NNLO fits including priors for the NNLO
LEC's, and then determine the NLO LEC's. 
Something like that approach has to be done here if comparisons with chiral perturbation theory are to be made.
It is, however, not so easy to do a fit to the complete NNLO formula for $m_{PS} a_0^{I=2}$:
the expression in Ref.~\cite{Bijnens:2011fm} contains 213 terms and involves a large number of LEC's.

\begin{figure}
\begin{center}
\includegraphics[width=0.8\textwidth,clip]{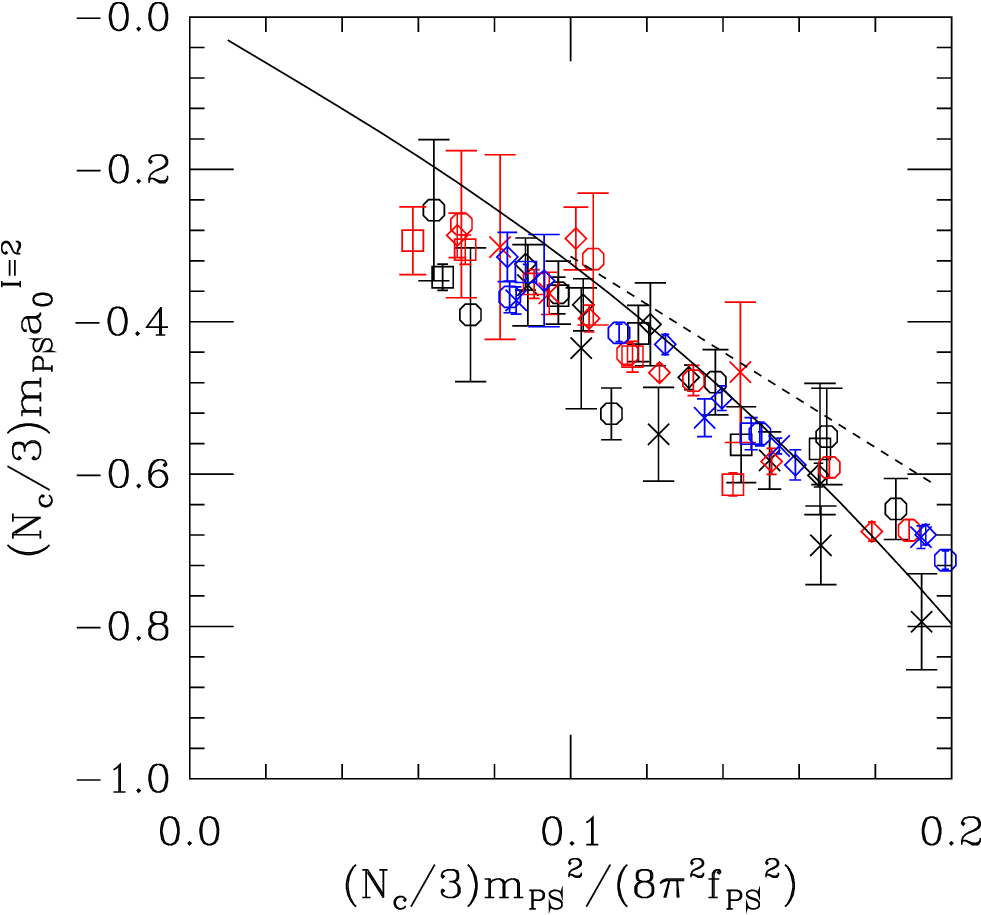}
\end{center}
\caption{ $(N_c/3)m_{PS}a_0^{I=2}$ versus  $\xi$. Black points are
$N_c=3$, red ones are $N_c=4$ and blue ones are $N_c=5$. Different plotting symbols label
different beta values and are listed in the text. The solid line is the curve of Eq.~\protect{\ref{eq:a0mps}}
with $l_{\pi\pi}=4.6$, the value of Ref.\protect{\cite{Feng:2009ij}}. The dotted line is the lowest order expression,
$m_{PS}a_0^{I=2} = -\pi \xi$.
\label{fig:ncmpia0vsncxi}}
\end{figure}

Let us look a bit more closely at the data.
The NLO chiral perturbation theory formula
for the $I=2$ scattering length $a_0^{I=2}$ is
\bee
a_0^{I=2} m_{PS} = -\pi \xi[ 1 - \frac{\xi}{2} (1 - 3 \ln(8\pi^2\xi) )- \frac{\xi}{2}l_{\pi\pi} ]
\label{eq:a0mps}
\ee
where $l_{\pi\pi}$ is a combination of low energy constants. Three pictures illustrate a
 comparison of Eq.~\ref{eq:a0mps} with my data:

Results for the quantity $(N_c/3)m_{PS}a_0^{I=2}$ as a function of $(N_c/3)\xi$ are presented 
in Fig.~\ref{fig:ncmpia0vsncxi}. The overall $N_c$ factor is redundant but it makes the axes of
Figs.~\ref{fig:ncmpia0vst0mpi2} and \ref{fig:ncmpia0vsncxi} look similar.
Different values of $N_c$ are again shown as different colors: black for $N_c=3$, red for $N_c=4$ and blue for $N_c=5$.
The different plotting symbols correspond to different bare gauge couplings; and are the same as in Fig.~\ref{fig:ncmpia0vst0mpi2}.
 The dotted line is the lowest order expression from
chiral perturbation theory, $m_{PS}a_0^{I=2} = -\pi \xi$. The solid line is the one loop perturbative 
formula  of Eq.~\ref{eq:a0mps}
with $l_{\pi\pi}=4.6$, the result from Ref.~\cite{Feng:2009ij}, also quoted by FLAG
\cite{FlavourLatticeAveragingGroupFLAG:2021npn}.

Next, I can divide out the leading term and show $-m_{PS}a_0^{I=2}/(\pi\xi)$
 versus  $(N_c/3)\xi$. The solid line is the curve of Eq.~\protect{\ref{eq:a0mps}}
with $l_{\pi\pi}=4.6$,  the value of Ref.~\protect{\cite{Feng:2009ij}}.

\begin{figure}
\begin{center}
\includegraphics[width=0.8\textwidth,clip]{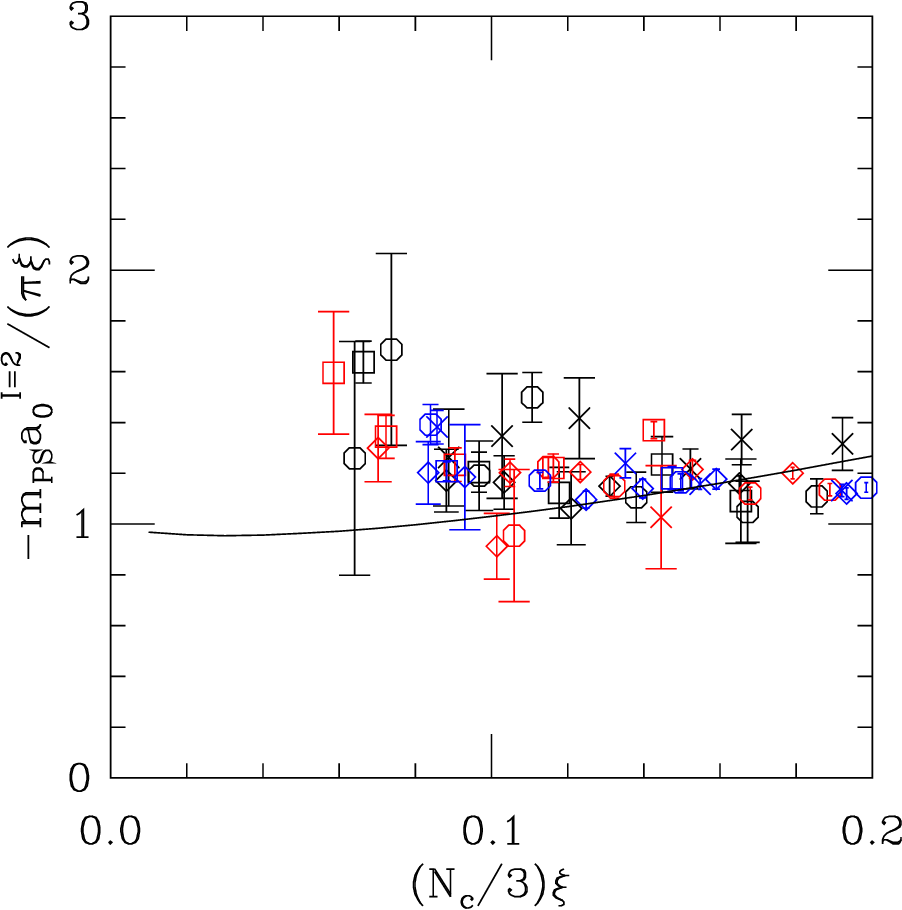}
\end{center}
\caption{ To expose deviations from the lowest order chiral perturbation theory formula, $-m_{PS}a_0^{I=2}/(\pi\xi)$
is plotted  versus  $(N_c/3)\xi$.  Black points are
$N_c=3$, red ones are $N_c=4$ and blue ones are $N_c=5$. Different plotting symbols label
different beta values and are 
the same as in Fig.~\protect{\ref{fig:ncmpia0vst0mpi2}} as listed in the text. The solid line is the curve of Eq.~\protect{\ref{eq:a0mps}}
with $l_{\pi\pi}=4.6$,  the value of Ref.~\protect{\cite{Feng:2009ij}}.
\label{fig:rvsncxi}}
\end{figure}

\begin{figure}
\begin{center}
\includegraphics[width=0.8\textwidth,clip]{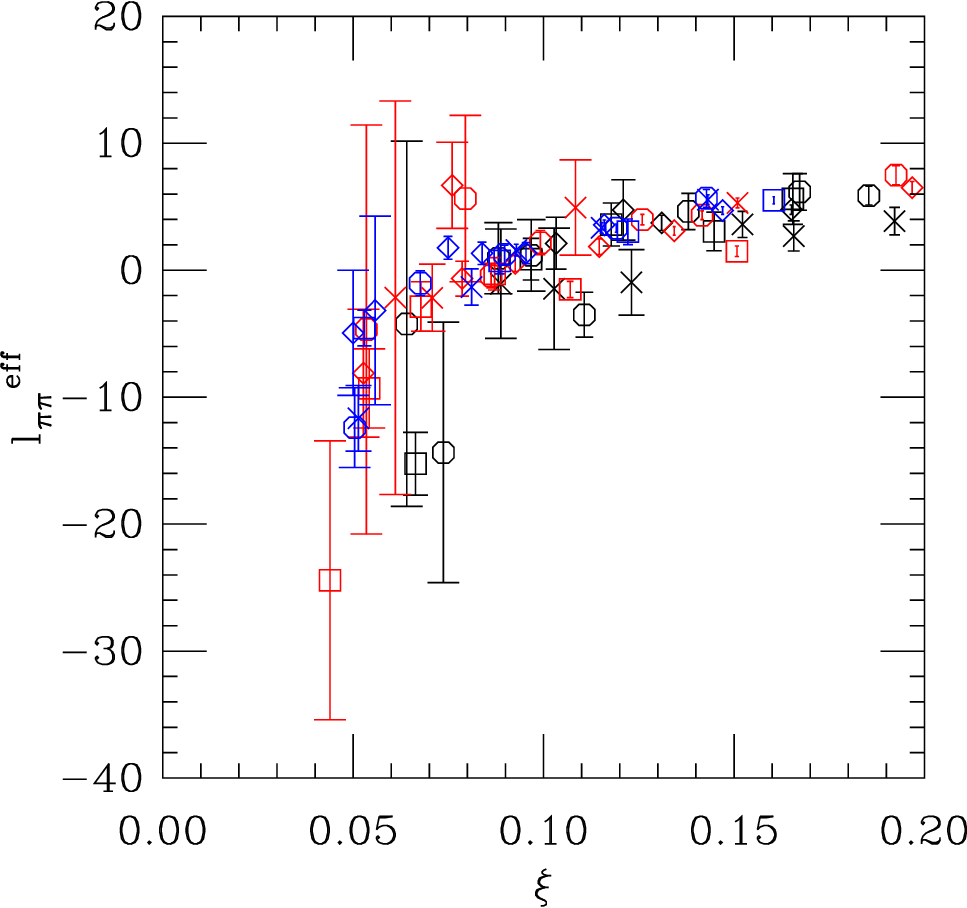}
\end{center}
\caption{ An ``effective''LEC $l_{\pi\pi}^{eff}$ versus $\xi$ from  inverting
Eq.~\protect{\ref{eq:a0mps}}: black points are
$N_c=3$, red ones are $N_c=4$ and blue ones are $N_c=5$.
Different plotting symbols label
different beta values and are listed in the text. 
\label{fig:lpipi}}
\end{figure}

Next, one might ask, is it possible to extract a value of $l_{\pi\pi}$ from the data? 
One can solve Eq.~\ref{eq:a0mps} for $l_{\pi\pi}$ as a function of $m_{PS}a_0$ and $\xi$ 
and plot that function -- an ``effective'' $l_{\pi\pi}^{eff}$ --  as it appears in the data. The  result of doing so
 is presented in Fig.~\ref{fig:lpipi}. If NLO chiral perturbation theory could describe the data, $l_{\pi\pi}^{eff}$ 
 would be a constant independent of $\xi$. 
 
This does not look promising! Besides the fact that the error bars are huge, the data
 for $l_{\pi\pi}$ are not a constant versus $\xi$, as the NLO formula requires.

But it is customary for lattice studies to include a fit of lattice data to some analytic formula.
So I tried, even though, from the point of a large $N_c$ comparison, it is an unnecessary exercise.
 Here is what I did:
The  expressions for the NNLO terms in Ref.~\cite{Bijnens:2011fm} come in four generic classes: 
\begin{enumerate}
\item
logarithms
 ( $\log m_{PS}^2)$ multiplied by analytic coefficients,
\item
squared logarithms $(\log m_{PS}^2)^2$ multiplied by analytic coefficients,
\item
logarithms multiplied by low energy constants,
 \item constant terms which are a mix of LEC's and analytic terms.
 \end{enumerate}
 (All of these are multiplied by a factor of $\xi^3$, of course). Based on the fact that the LEC's
(in Bijnen's and Lu's conventions) are small numbers, I keep the squared logarithms and logarithms with their analytic
coefficients, and include a constant term whose value is a free parameter to be fit. Translating conventions,
I assume that the NNLO term is
\bee
(m_{PS}a_0^{I=2})_{NNLO} = - \pi \xi \left[  \xi^2(A_2 L^2 + A_1 L + C) \right]
\label{eq:nnlo}
\ee
where $A_2=31/6$, $A_1=107/9$, $L= \log(8\pi^2\xi)$ and $C$ is the free parameter.

 We expect there to be
lattice artifacts, so (as in Ref.~\cite{DeGrand:2023hzz}) I added an overall nuisance parameter to give
\bee
m_{PS}a_0^{I=2} = (1+ c_a \frac{a^2}{t_0})((m_{PS}a_0^{I=2})_{NLO} + (m_{PS}a_0^{I=2})_{NNLO} ) .
\label{eq:nnlo_model}
\ee
(The NLO term in given in Eq.~\ref{eq:a0mps}.)
It turns out that the data are noisy enough that fits are insensitive to this addition, and it was not kept in
the results shown in what follows.

\begin{figure}
\begin{center}
\includegraphics[width=0.8\textwidth,clip]{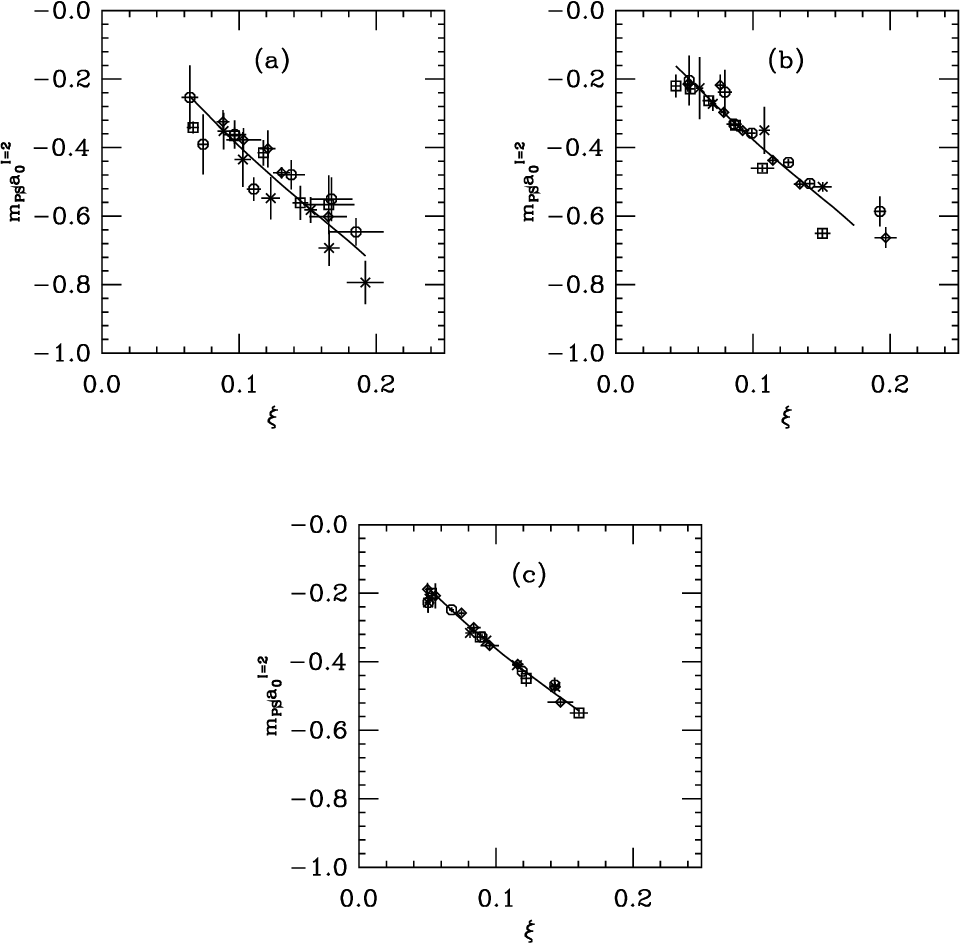}
\end{center}
\caption{ Data for $m_{PS}a_0^{I=2}$ versus $\xi$ with curves showing
 a fit to the approximate NNLO formula, the sum of Eqs.~\protect{\ref{eq:a0mps}} and \protect{\ref{eq:nnlo}}.
Panels (a), (b) and (c) are for $N_c=3$, 4, and 5, respectively.
 The solid lines
are the  result of the fit function.
\label{fig:fit3parall}}
\end{figure}

Both the ``$y$'' ($m_{PS}a_0^{I=2}$) and ``$x$'' ($\xi$) variables have uncertainties.
I deal with the latter ones as in Ref.~\cite{DeGrand:2023hzz}, by adding a fit parameter $\xi_{fit}(i)$
for each $\xi_i$ (for bare parameter set $i$) and augmenting the chi-squared formula with
an extra $\sum_i (\xi_{fit}(i)-\xi_i)^2/\sigma(\xi_i)^2$ term.
Fits to the model perform reasonably well, at least from the point of view of a chi-squared value.
The fit parameters are poorly determined.  Fitted values of $l_{\pi\pi}$ do not match
the $SU(3)$ value of Ref.~\cite{Feng:2009ij}.
For completeness, a picture of a fit to all the data for each value of $N_c$ is shown
 in Fig.~\ref{fig:fit3parall}, and fit results from those fits are collected in  Table ~\ref{tab:fit3parall}.
More reliable results come from model averaging over each $N_c$, keeping at least ten bare parameter values in the
average. These results are shown in Table \ref{tab:fit3modave}.
There is weak (but monotomic) dependence of the fit parameters on $N_c$.
 (Recall that one expects $l_{\pi\pi} \sim a+bN_c$, $C \sim d+e N_c+ f N_c^2$.)

A comparison of Figs.~\ref{fig:rvsncxi} and \ref{fig:fit3parall} shows that the analytic terms are fighting with
the chiral logarithms to  flatten the dependence of $m_{PS} a_0^{I=2}$ on $\xi$.

Finally, at large $N_c$ the eta prime mass  is expected to fall and to contribute to the chiral expansion.
The authors of Ref.~\cite{Baeza-Ballesteros:2022azb} have done the calculation of the scattering
length in this so-called $U(N_f)$ case and in NNLO they find
\bee
a_0^{I=2} m_{PS} = -\pi \xi[ A + B \xi\ln(\frac{m_{PS}^2}{\mu^2}) + C\xi \ln(\frac{m_{\eta'}^2}{\mu^2} ]
\label{eq:a0mpslarge}
\ee
where for $N_f=2$ the coefficients are
\beea
A &=& 1 - a_1\xi  + a_2 \xi^2\nonumber \\
B &=& \frac{3}{2}\nonumber \\
C &=& \frac{1}{2} . \nonumber \\
\label{eq:abc}
\eea
To convert to the language of Ref.~\cite{Baeza-Ballesteros:2022azb}, $a_1=(16\pi)^2L_{SS}$ and $a_2= (16\pi^2)^2 K_{SS}$.
Ref.~\cite{Baeza-Ballesteros:2022azb} points out  that $a_1$ should scale as $N_c$ and
$a_2$ should scale as $N_c^2$.

The eta prime mass is given by the Witten-Veneziano   \cite{Witten:1979vv,Veneziano:1979ec}  formula
and enters in the expression through $\xi_\eta= M_\eta^2/(8\pi^2F_\pi^2)$ where
\bee
M_{\eta'}^2 = M^2 + M_0^2
\label{eq:etaprime}
\ee
where the zero-quark mass eta prime mass is
\bee
M_0^2 =  \frac{4 N_f \chi_T}{F^2}.
\label{eq:meta}
\ee
and $\chi_T$ is the quenched topological susceptibility.
Ref.~\cite{Ce:2016awn} measured $\chi_T$ to be
\bee
t_0^2 \chi_T= 7\times 10^{-4}.
\ee
 (See also \cite{Ce:2015qha} for earlier determinations of $\chi_T$.)
 
 I attempted a fit to Eq.~\ref{eq:a0mpslarge} with the following choices: Following Ref.~\cite{DeGrand:2023hzz} (and FLAG fits to 
 $m_{PS}^2$ and $f_{PS}$) I took $\mu^2$ equal to the physical squared pion mass. I included values of
 $t_0 m_{PS}^2$ and $\sqrt{t_0}f_{PS}$ in the fit (rather than a common $\xi$ value) and kept the correlations between them.
 $t_0^2 \chi_T$ was fixed, but $M_{\eta'}^2 $ depended on $t_0 m_{PS}^2$ and $\sqrt{t_0}f_{PS}$.
 (Including  lattice spacing dependence with a nuisance parameter,
 multiplying the right hand side of Eq.~\ref{eq:a0mpslarge}
 by $(1+c_a(a^2/t_0))$ as in Eq.~\ref{eq:nnlo_model}, did not affect the fits and it was not kept.)
 For $N_D$ data points, this is a fit of $3N_D$ quantities ($t_0 m_{PS}^2$, $\sqrt{t_0}f_{PS}$, and $m_{PS}a_0$ per point)
 to $2N_D + 2$ parameters (fitted $t_0 m_{PS}^2$, $\sqrt{t_0}f_{PS}$ and $m_{PS}a_0$ per point, plus $a_1$
and  $a_2$).

Results of these fits were  unsatisfactory. Fits to
individual $\beta$ sets (without $c_a$) could be achieved, but the values of $a_1$ and $a_2$ 
had large uncertainites. One example is shown in Table ~\ref{tab:fit3u}.
 
I think that there is not much to be gained by performing yet more elaborate fits to the data.

\section{Conclusions\label{sec:conc}}

Fig.~\ref{fig:ncmpia0vsncxi} shows that once again lattice studies 
of a nonperturbative quantity
agree qualitatively with large $N_c$ expectations. In the $I=2$ channel, pions scattering at threshold
interact less and less strongly as $N_c$ increases.  The scaling law that $m_{PS}a_0^{I=2}$ is 
proportional to $1/N_c$ was observed,
 across the range of pion masses studied.

\begin{table}
\begin{tabular}{c c c c c c c c}
\hline
$\beta$ & $\kappa$ & $L$ & $\xi$ & $\Delta E$ & $m_{PS}a_0^{I=2}$ & $am_{PS}$  & $t_0/a^2$ \\
\hline
5.25 & 0.1284 & 16 & 0.165(19) & 0.0096(18) & -0.567(86) & 0.475(2)  & 1.037 \\ 
5.25 & 0.1288 & 16 & 0.145(6) & 0.0129(15) & -0.561(50) & 0.414(3)  & 1.037 \\ 
5.25 & 0.1292 & 16 & 0.118(4) & 0.0141(16) & -0.415(37) & 0.337(3)  & 1.037 \\ 
5.25 & 0.1294 & 16 & 0.097(5) & 0.0168(13) & -0.366(24) & 0.290(3)  & 1.037 \\ 
5.25 & 0.1296 & 16 & 0.066(4) & 0.0219(15) & -0.342(17) & 0.248(3)  & 1.037 \\ 
\hline
5.3 & 0.1280 & 16 & 0.165(14) & 0.0158(6) & -0.602(16) & 0.394(3)  & 1.374 \\ 
5.3 & 0.1284 & 16 & 0.131(6) & 0.0182(9) & -0.473(16) & 0.323(3)  & 1.374 \\ 
5.3 & 0.1285 & 16 & 0.121(6) & 0.0173(29) & -0.403(54) & 0.302(3)  & 1.374 \\ 
5.3 & 0.1286 & 16 & 0.103(13) & 0.0221(10) & -0.378(34) & 0.262(13)  & 1.374 \\ 
5.3 & 0.1288 & 24 & 0.088(5) & 0.0064(8) & -0.324(34) & 0.234(2)  & 1.374 \\ 
\hline
5.35 & 0.1270 & 16 & 0.185(20) & 0.0156(12) & -0.646(40) & 0.413(7)  & 1.804 \\ 
5.35 & 0.1275 & 16 & 0.167(16) & 0.0202(28) & -0.551(63) & 0.337(5)  & 1.804 \\ 
5.35 & 0.1278 & 16 & 0.138(10) & 0.0204(24) & -0.479(43) & 0.310(4)  & 1.804 \\ 
5.35 & 0.1280 & 24 & 0.111(5) & 0.0094(8) & -0.521(34) & 0.255(3)  & 1.804 \\ 
5.35 & 0.1282 & 24 & 0.097(8) & 0.0083(12) & -0.362(41) & 0.220(2)  & 1.804 \\ 
5.35 & 0.1283 & 24 & 0.074(4) & 0.0125(35) & -0.391(88) & 0.192(2)  & 1.804 \\ 
5.35 & 0.1284 & 24 & 0.064(6) & 0.0098(42) & -0.253(93) & 0.168(3)  & 1.804 \\ 
\hline
5.4 & 0.1265 & 24 & 0.192(13) & 0.0061(6) & -0.794(63) & 0.390(3)  & 2.379 \\ 
5.4 & 0.1270 & 24 & 0.166(8) & 0.0077(8) & -0.693(52) & 0.326(2)  & 2.379 \\ 
5.4 & 0.1272 & 24 & 0.152(5) & 0.0077(6) & -0.582(38) & 0.296(2)  & 2.379 \\ 
5.4 & 0.1276 & 24 & 0.123(7) & 0.0119(17) & -0.548(62) & 0.236(5)  & 2.379 \\ 
5.4 & 0.1277 & 24 & 0.103(6) & 0.0113(27) & -0.435(79) & 0.212(2)  & 2.379 \\ 
5.4 & 0.1278 & 24 & 0.089(4) & 0.0119(22) & -0.352(53) & 0.185(2)  & 2.379 \\ 
\hline
\end{tabular}
\caption{Data for $N_c=3$. Values of $t_0$ from  Ref.~\cite{DeGrand:2023hzz} have been appended.
\label{tab:su3tab}}
\end{table}

\begin{table}
\begin{tabular}{c c c c c c c c}
\hline
$\beta$ & $\kappa$ & $L$ & $\xi$ & $\Delta E$ & $m_{PS}a_0^{I=2}$ & $am_{PS}$  & $t_0/a^2$ \\
\hline
10.0 & 0.1270 & 16 & 0.151(6) & 0.0063(2) & -0.650(15) & 0.618(1)  & 1.116 \\ 
10.0 & 0.1280 & 16 & 0.107(8) & 0.0085(3) & -0.460(11) & 0.448(4)  & 1.116 \\ 
10.0 & 0.1285 & 16 & 0.087(4) & 0.0103(6) & -0.334(15) & 0.345(2)  & 1.116 \\ 
10.0 & 0.1288 & 16 & 0.068(5) & 0.0122(6) & -0.263(14) & 0.280(5)  & 1.116 \\ 
10.0 & 0.1289 & 16 & 0.054(3) & 0.0129(8) & -0.229(14) & 0.253(5)  & 1.116 \\ 
10.0 & 0.1290 & 16 & 0.044(3) & 0.0154(28) & -0.220(33) & 0.229(4)  & 1.116 \\ 
\hline
10.1 & 0.1250 & 16 & 0.197(8) & 0.0063(4) & -0.663(30) & 0.624(2)  & 1.770 \\ 
10.1 & 0.1266 & 16 & 0.134(5) & 0.0103(3) & -0.506(10) & 0.433(1)  & 1.770 \\ 
10.1 & 0.1270 & 16 & 0.115(4) & 0.0118(4) & -0.437(13) & 0.376(4)  & 1.770 \\ 
10.1 & 0.1275 & 16 & 0.093(5) & 0.0149(4) & -0.350(9) & 0.300(2)  & 1.770 \\ 
10.1 & 0.1277 & 16 & 0.079(2) & 0.0147(9) & -0.297(13) & 0.275(2)  & 1.770 \\ 
10.1 & 0.1278 & 16 & 0.076(3) & 0.0121(20) & -0.218(31) & 0.255(2)  & 1.770 \\ 
10.1 & 0.1280 & 16 & 0.053(4) & 0.0196(19) & -0.215(22) & 0.202(5)  & 1.770 \\ 
\hline
10.2 & 0.1252 & 16 & 0.193(5) & 0.0095(9) & -0.586(43) & 0.487(2)  & 2.503 \\ 
10.2 & 0.1262 & 16 & 0.142(3) & 0.0141(4) & -0.505(11) & 0.376(2)  & 2.503 \\ 
10.2 & 0.1265 & 16 & 0.126(3) & 0.0149(4) & -0.443(10) & 0.341(2)  & 2.503 \\ 
10.2 & 0.1270 & 16 & 0.099(3) & 0.0190(10) & -0.358(15) & 0.272(2)  & 2.503 \\ 
10.2 & 0.1272 & 16 & 0.086(5) & 0.0231(8) & -0.332(11) & 0.239(3)  & 2.503 \\ 
10.2 & 0.1273 & 16 & 0.079(5) & 0.0183(60) & -0.238(65) & 0.221(3)  & 2.503 \\ 
10.2 & 0.1275 & 24 & 0.053(2) & 0.0070(28) & -0.204(72) & 0.175(1)  & 2.503 \\ 
\hline
10.3 & 0.1260 & 16 & 0.151(6) & 0.0183(6) & -0.515(14) & 0.339(2)  & 3.339 \\ 
10.3 & 0.12675 & 16 & 0.108(4) & 0.0233(58) & -0.350(69) & 0.245(2)  & 3.339 \\ 
10.3 & 0.1271 & 24 & 0.071(2) & 0.0093(8) & -0.272(21) & 0.179(2)  & 3.339 \\ 
10.3 & 0.1272 & 24 & 0.061(3) & 0.0096(45) & -0.226(91) & 0.159(2)  & 3.339 \\ 
\hline
\end{tabular}
\caption{Data for $N_c=4$.
\label{tab:su4tab}}
\end{table}

\begin{table}
\begin{tabular}{c c c c c c c c}
\hline
$\beta$ & $\kappa$ & $L$ & $\xi$ & $\Delta E$ & $m_{PS}a_0^{I=2}$ & $am_{PS}$  & $t_0/a^2$ \\
\hline
16.2 & 0.1250 & 16 & 0.160(7) & 0.0047(1) & -0.549(12) & 0.646(2)  & 1.510 \\ 
16.2 & 0.1260 & 16 & 0.122(4) & 0.0057(3) & -0.449(24) & 0.530(2)  & 1.510 \\ 
16.2 & 0.1270 & 16 & 0.088(5) & 0.0073(3) & -0.328(13) & 0.399(2)  & 1.510 \\ 
16.2 & 0.1278 & 16 & 0.053(2) & 0.0096(4) & -0.201(6) & 0.272(2)  & 1.510 \\ 
\hline
16.3 & 0.1250 & 16 & 0.147(9) & 0.0065(2) & -0.518(10) & 0.538(3)  & 1.959 \\ 
16.3 & 0.1260 & 16 & 0.116(4) & 0.0077(2) & -0.408(8) & 0.438(2)  & 1.959 \\ 
16.3 & 0.1264 & 16 & 0.095(7) & 0.0085(3) & -0.353(12) & 0.388(1)  & 1.959 \\ 
16.3 & 0.1268 & 16 & 0.084(5) & 0.0098(4) & -0.300(10) & 0.333(1)  & 1.959 \\ 
16.3 & 0.1270 & 16 & 0.075(2) & 0.0099(4) & -0.258(8) & 0.306(1)  & 1.959 \\ 
16.3 & 0.1273 & 16 & 0.056(2) & 0.0113(23) & -0.208(36) & 0.256(1)  & 1.959 \\ 
16.3 & 0.1275 & 16 & 0.050(2) & 0.0137(17) & -0.189(19) & 0.222(2)  & 1.959 \\ 
\hline
16.4 & 0.1252 & 16 & 0.143(3) & 0.0078(5) & -0.468(22) & 0.469(2)  & 2.468 \\ 
16.4 & 0.1258 & 16 & 0.119(3) & 0.0099(2) & -0.428(7) & 0.403(2)  & 2.468 \\ 
16.4 & 0.1265 & 16 & 0.090(3) & 0.0116(4) & -0.328(9) & 0.323(2)  & 2.468 \\ 
16.4 & 0.1270 & 16 & 0.068(2) & 0.0152(4) & -0.249(7) & 0.246(3)  & 2.468 \\ 
16.4 & 0.1272 & 24 & 0.050(2) & 0.0054(8) & -0.227(29) & 0.211(1)  & 2.468 \\ 
\hline
16.6 & 0.1252 & 16 & 0.143(4) & 0.0111(4) & -0.473(13) & 0.403(1)  & 3.487 \\ 
16.6 & 0.1260 & 16 & 0.115(3) & 0.0159(4) & -0.410(9) & 0.317(2)  & 3.487 \\ 
16.6 & 0.1264 & 16 & 0.093(2) & 0.0181(4) & -0.338(6) & 0.269(2)  & 3.487 \\ 
16.6 & 0.1266 & 16 & 0.081(3) & 0.0210(5) & -0.316(15) & 0.242(6)  & 3.487 \\ 
16.6 & 0.1269 & 24 & 0.051(1) & 0.0071(4) & -0.216(10) & 0.182(1)  & 3.487 \\ 
\hline
\end{tabular}
\caption{Data for $N_c=5$.
\label{tab:su5tab}}
\end{table}

\begin{table}
\begin{tabular}{c | c c c}
\hline
$N_c$ & $l_{\pi\pi}$ & $C$ & $\chi^2/DoF$\\
\hline
3 & $-15.5 \pm 1.9$ & $-7.8\pm 0.4$  & $28.4/20$ \\
4 & $-13.4 \pm 1.0$ & $-8.4 \pm 0.2$  & $45.1/22$  \\
5 & $-12.4 \pm 1.0$ & $-9.1 \pm 2.4$  & $10.6/19$ \\
\hline
\end{tabular}
\caption{Results from fits to Eq.~\ref{eq:nnlo_model} to all the data at each $N_c$ value.
\label{tab:fit3parall}}
\end{table}

\begin{table}
\begin{tabular}{c | c c }
\hline
$N_c$ & $l_{\pi\pi}$ & $C$  \\
\hline
3 & $-16.8 \pm 2.6$ & $-8.1\pm 0.6$  \\
4 & $-14.3 \pm 3.9$ & $-8.7 \pm 1.4$    \\
5 & $-13.3 \pm 1.3$ & $-9.3 \pm 0.3$   \\
\hline
\end{tabular}
\caption{Model averaged parameter values from fits to Eq.~\ref{eq:nnlo_model}.
\label{tab:fit3modave}}
\end{table}

\begin{table}
\begin{tabular}{c | c c c}
\hline
$N_c$ & $a_1$ & $a_2$ & $\chi^2/DoF$\\
\hline
3 & $-1.6 \pm 0.9$ & $-42.0\pm 5.4$  & $36/20$ \\
4 & $0.05 \pm 0.4$ & $-39.7 \pm 2.9$  & $60/22$  \\
5 & $1.5 \pm 0.4$ & $-48.4 \pm 2.9$  & $38/19$ \\
\hline
\end{tabular}
\caption{Results from fits to Eq.~\ref{eq:a0mpslarge}.
\label{tab:fit3u}}
\end{table}

\begin{acknowledgments}
My computer code is based on the publicly available package of the
 MILC collaboration~\cite{MILC}. The version I use was originally developed by Y.~Shamir and
 B.~Svetitsky.
This material is based upon work supported by the U.S. Department of Energy, Office of Science, Office of
High Energy Physics under Award Number DE-SC-0010005.
Some of the computations for this work were also carried out with resources provided by the USQCD
Collaboration, which is funded
by the Office of Science of the U.S.\ Department of Energy
using the resources of the Fermi National Accelerator Laboratory (Fermilab), a U.S.
Department of Energy, Office of Science, HEP User Facility. Fermilab is managed by
 Fermi Research Alliance, LLC (FRA), acting under Contract No. DE- AC02-07CH11359.
\end{acknowledgments}

%%%%%%%%%%%%%%%%%%%%%%%%%%%%%%%%%%%%%%%%%%%%%%%%%%%%%%%%%%%%%%%%%%%%%

%%%%%%%%%%%%%%%%%%%%%%%%%%%%%%%%%%%%%%%%%%%%%%%%%%%%%%%%%%%%%%%%%%%%%

\end{document}